\documentclass[12pt]{article} 

\usepackage[utf8]{inputenc}
\usepackage[margin=0.65in]{geometry}
\usepackage{setspace}                
\usepackage{titlesec}                
\usepackage{hyperref}

\usepackage{authblk} 
\usepackage{graphicx}      
\usepackage{subfig}       
\usepackage{caption}      
\usepackage[numbers,sort&compress]{natbib}

\titlespacing*{\section}{0pt}{1.5ex plus 1ex minus .2ex}{1ex plus .2ex}

\title{\textbf{Analyzing Interaction Between CCAs and Traffic Policers}}

\author{Ammar Tahir \\ UIUC \\ 
Champaign, IL \\ \textit{ammart2@illinois.edu}}

\date{} 

\begin{document}

\maketitle
\thispagestyle{empty}

\begin{abstract}
    We describe details of a formal framework to study the interaction between traffic policers, implemented using phantom queues or token buckets, and any arbitrary congestion control algorithm (CCA). This framework allows network providers to figure out configurations for their traffic policers (phantom queue size, safe rate thresholds, etc.). We also use this framework to describe why CCAs interact differently with a traffic policer compared to traffic shapers.
\end{abstract}

\section{Introduction}

Traffic policers are a well-known and widely used mechanism to implement rate-limiting in modern networks. Compared to traffic shapers, the other common mechanism for rate-limiting, traffic policers are cheap and lightweight, as they do not need to store packets in memory buffers. Despite this, traffic policers are considered an inferior mechanism, as they interact poorly with widely deployed congestion control algorithms (like TCP Cubic, Reno, and BBR), often leading to high bursts, poorly enforced rates, and high drop rates. Our recent wor, BC-PQP~\cite{bcpqp}, has proposed mechanisms to improve different aspects of traffic policers: better average rate conformity, controlled bursts, and the ability to implement various rate-sharing policies within the rate-limited traffic aggregate. This short write up builds on insights from BC-PQP~\cite{bcpqp} to formally study the interaction between any arbitrary CCA and a traffic policer. Our analysis not only helps explain why CCAs interact quite differently with a traffic policer compared to a traffic shaper, but also helps operators configure traffic policers better. 
\section{Framework}

We are going to model an arbitrary CCA using a fluid model; thus, the sending rate of the CCA at time $t$ is given by $r(t)$. 
Given that traffic policers lack physical buffers, our flow does not incur any queuing delay and only experiences packet losses as a signal of congestion (phantom queue being full).
Thus, we model a CCA with two functions: 1) an increment function, $Inc$, which increases the sending rate $r(t)$ over time, i.e., $\frac{dr}{dt}$, represented as, $\dot{r}(t)$, from now onwards, and 2) a decrement function, $Dec(r)$, which decreases the sending rate $r(t)$ after a loss event.

We now assume that CCA is in its stable stage, i.e., it has exited its initial slow-start phase, and the entire CCA behavior is modeled by the $Inc = \dot{r}(t)$ and $Dec(r)$ functions. In this stage, we call $r_h$ to be the rate right before a packet loss event (and accompanying rate reduction) and $r_l$ the rate right after a loss event. Then we have:

\begin{equation} \label{eq:1}
r_l = Dec(r_h)
\end{equation}

After the rate reduction, CCA increments its rate from $r_l$ over time. At any time $\tau$ since this rate reduction, rate, $r(\tau)$, can be given as:

\begin{equation} \label{eq:2}
r(\tau) = r_l + \int_0^{\tau} \dot{r}(t) dt
\end{equation}

Furthermore, we can find bytes sent over this duration by integrating $r(\tau)$ function w.r.t time $t$:

\begin{equation} \label{eq:3}
A(\tau) = \int_0^{\tau} r(t) dt
\end{equation}

This can be expanded and simplified as follows:

$$A(\tau) = r_l \tau + \int_0^{\tau}  \int_0^{u} \dot{r}(u) dudt$$

\begin{equation} \label{eq:A}
A(\tau) = r_l \tau + \int_0^{\tau} (\tau - t)\dot{r}(t) dt
\end{equation}

\subsection{Sizing the Traffic Policer}

We now describe how to find the phantom queue size (or interchangably token bucket size) that ensures that any flow using the defined CCA can maintain set service rate $r$ on average over time. Note that after a rate decrement, triggered by a packet loss once phantom queue became fill, the rate is reduced to $r_l < r$. If it takes $\tau_l$ time for the rate to go from $r_l$ to $r$, the flow has a byte deficit of $A(\tau_l) - r\tau_l$. Thus phantom queue should be sized at least greater than this to ensure that flow is able to send deficit amount of bytes once its rate starts exceeding service rate $r$.

To calculate this, we first compute $\tau_l$ by putting $r$ on the right hand side of Eq.~\ref{eq:2}. Then we calculate phantom queue size $Q$ by the following relationship:

\begin{equation} \label{eq:Q}
Q \geq \max (0, r\tau_l - A(\tau_l))
\end{equation}

Note that we need to know the relationship between $r_l$ and $r$ to calculate the above relationship, which we show how to calculate in the next section.

\subsection{Safe Rate Thresholds}

As discussed in \cite{bcpqp} and shown later, for most CCAs the queue size calculated with Eq.~\ref{eq:Q} is often quite large. While this sizing is needed for correct rate enforcement for a flow in steady state, during a flow's slow start phase when it is increasing the sending rate exponentially (often in the form of doubling congestion window), we can end up with spuriously high sending rates before flow incurs first packet drop. It takes a high number of packet drops and resultant rate reductions before the flow is able to converge to correct rate. To avoid this, BC-PQP monitors the sending rate and if the flow exceeds `safe rate thresholds`, it fills up the phantom queue to incur packet drops and put the flow in its steady state. We now discuss how to calculate these safe rate thresholds. Specifically, we will calculate $r_l$ and $r_h$, the safe rate threshold should be set lower and higher than these respecitvely. We specifically find time, $\tau_h$, taken to reach rate $r_h$ starting from $r_l$ by plugging $r_h$ on right hand side of Eq.~\ref{eq:2} with $r_l = Dec(r_h)$. 
We can then use this $\tau_h$ in Eq.\ref{eq:A} with $A(\tau_h) = r \tau_h$, i.e.,:

\begin{equation} \label{eq:5}
r \tau_h = r_l \tau_h + \int_0^{\tau_h} (\tau_h - t)\dot{r}(t) dt
\end{equation}

Note that $\tau_h$ cancels out on both sides after solving the integral, leaving us with a simple relationship between $r$, $r_l$, and $r_h$, which can be solved using Eq.\ref {eq:1} to find relationship between $r$ and either of $r_l$ and $r_h$.

\subsection{Congestion Window Based CCAs}

Several CCAs, like TCP Reno and TCP Cubic, operate using a congestion window instead of a sending rate. They rely on the ack-clocking property of TCP to do rate control. If $\omega(t)$ is the congestion window and $R(t)$ is the round-trip time of the flow at time $t$, we can convert window to rate as follows:

\begin{equation} \label{eq:r_w}
r(t) = \frac{\omega(t)}{R(t)}
\end{equation}

Then we can find $\dot{r}(t)$ as:

\begin{equation} \label{eq:r_dot}
\dot{r}(t) = \frac{\frac{d\omega}{dt} R(t) - \frac{dR}{dt} \omega(t)}{R^2(t)}
\end{equation}

We remind ourselves that RTT does not change over time with a traffic policer, and is just equal to the propagation delay, $D$. This, as we discuss later, is a critical difference in how such CCAs interact with policers versus shapers. However, for a policer, this can be simplified as:

\begin{equation} \label{eq:r_dot2}
\dot{r}(t) = \frac{d\omega}{dt} \frac{1}{D}
\end{equation}

We will now apply this framework to a few CCAs.

\section{CCA Examples}

We will discuss a few examples to show how this framework can be used for different CCAs.

\subsection{TCP Reno}

TCP Reno is a window-based CCA that increments its congestion window by 1 MSS packet on the successful delivery of all packets from the previous window, i.e., it increments the congestion window by 1 each round-trip cycle, $C$. That is, $\frac{d\omega}{dC} = 1$, also not that $\frac{dC}{dt} = 1 / R(t)$, and we also know that for a policer RTT, $R(t)$, is just a fixed propagation delay $D$. Therefore, we have:

$$\frac{d\omega}{dt} = \frac{d\omega}{dC} \cdot \frac{dC}{dt} = \frac{1}{D}$$

Putting this in Eq.~\ref{eq:r_dot2}, we get the following $Inc$ function for TCP Reno:

$$Inc_{Reno} = \dot{r}(t) = \frac{1}{D^2}$$

On the other hand, $Dec_{Reno}$ function is simply $Dec(r) = \beta r$ with $\beta = 0.5$. Now we can apply our framework to find queue size and rate thresholds. 
We begin by finding the rate thresholds, $r_l$ and $r_h$.
First, we calculate the time taken to go from $r_l$ to $r_h$, i.e., $\tau_h$:

$$r_h = \frac{1}{2} r_h + \int_0^{\tau_h} \frac{1}{D^2} dt$$

\begin{equation} \label{eq:10}
\tau_h = \frac{1}{2} r_h D^2
\end{equation}

When we plug $\dot{r}(t)$ for Reno in Eq.~\ref{eq:5}, we get:

$$r \tau_h = r_l \tau_h + \int_0^{\tau_h} (\tau_h - t)\frac{1}{D^2} dt$$
$$r = r_l + \frac{\tau_h}{2D^2}$$

Plugging the value of $\tau_h$ from above alongside the relationship between $r_h$ and $r_l$ from $Dec$ function, we get:

$$r = \frac{1}{2} r_h + \frac{1}{4} r_h$$

\begin{equation} \label{eq:r_reno}
r_h = \frac{4}{3} r, r_l = \frac{2}{3} r
\end{equation}

Now, we can calculate the phantom queue size $Q$ of the policer using Eq.~\ref{eq:Q}. We start by finding the time taken, $\tau_l$, to go from $r_l$ to $r$, the policed rate. Since we also know $r_l$ in terms of $r$, we get the following using Eq.~\ref{eq:2}:

$$r = \frac{2}{3} r + \frac{\tau_l}{D^2}$$
$$\tau_l = \frac{1}{3} rD^2$$

Now we can find $A(\tau_l)$ using Eq.~\ref{eq:A}:

$$A(\tau_l) = \frac{5}{18} r^2D^2$$

Using Eq.~\ref{eq:Q}, we get the following rule for sizing a queue for TCP Reno flow:

$$Q \geq \frac{1}{18} r^2 D^2$$

\subsection{TCP Cubic}

TCP Cubic is another window-based protocol; however, it uses a cubic polynomial function for the window increments. Specifically, it increments its congestion window based on the time elapsed, $t$, since the last window reduction using the following formula: 

$$\omega(t) = \omega_{max} + C(t - K)^3$$

Where $K = \sqrt[3]{\omega_{max} \frac{1 - \beta}{C}}$, and $\beta = 0.7$ and $C = 0.4$. Then $\frac{d\omega}{dt} = 3C(t - K)^2$ and putting this in Eq.~\ref{eq:r_dot2}, we get $Inc_{cubic}$:

$$Inc_{cubic} = \dot{r}(t) = \frac{3C(t-K)^2}{D}$$

On the other hand, $Dec_{cubic}(r) = \beta r$. We again begin by finding $\tau_h$ to eventually find safe rate thresholds. Using the observation that $\omega_{max} = r_hD$, we find safe rate thresholds to be following:







$$r_h = \frac{4}{3 + \beta}r, r_l = \frac{4 \beta}{3 + \beta}r$$

With $\beta = 0.7$, this evaluates to:

$$r_h = \frac{40}{37}r, r_l = \frac{28}{37}r$$

Using the above, we evaluate the phantom queue size needed to do correct rate enforcement for QLDL, which evaluates to:

$$Q \geq \frac{3}{4\sqrt[3]{C}} (\frac{1 - \beta}{3 + \beta})^{\frac{4}{3}} r \sqrt[3]{rD}$$

Plugging $\beta$ and $C$, we get the following with $G \approx 0.03572$:

$$Q \geq G r \sqrt[3]{rD}$$

\subsection{GCC}

GCC is a rate-based protocol used in WebRTC. It consists of both a sender-side loss-based component and a receiver-side delay-based component. Since the delay does not change with a policer, we only need to model the loss-based part of GCC. The sender increments its rate by a multiplicative increase of $\alpha = 1.05\times$ the last sending rate every time it receives an RTCP report from the receiver (sent every $\delta$ seconds) with losses fewer than 2\%. Thus, we can define the $Inc_{GCC}$ as follows:

$$Inc_{GCC} = \dot{r}(t) = Kr_le^{Kt}$$

Here, $r_l$ is the rate right after a reduction post packet loss event, and $K =\frac{\ln(\alpha)}{\delta}$. Moreover, $Dec$ function for GCC depends on the fraction of lost packets, specifically, if the fraction of lost packets exceeds 20\%, the rate is reduced by a multiplicative factor $\beta = 1 - \frac{Loss}{2}$. 
We apply our framework to find rate thresholds. To do this, we first find $\tau_h$:

$$r_h = r_l + \int_0^{\tau_h} Kr_le^{Kt} dt$$
$$r_h = r_l + r_le^{K\tau_h} - r_l$$
$$\tau_h = \delta \log_{\alpha}\frac{1}{\beta}$$

Using this in Eq.\ref{eq:5}, we get:

$$r \tau_h = r_l \tau_h + \int_0^{\tau_h} (\tau_h - t)Kr_le^{Kt} dt$$

$$r \tau_h = \frac{r_l(e^{K\tau_h} -1)}{K}$$
$$r_l = \frac{-\beta r \ln(\beta)}{1 - \beta}, r_h = \frac{-r \ln(\beta)}{1 - \beta}$$

Using the above, we can now calculate the phantom queue size needed to do correct rate enforcement, which comes out to be:

$$Q \geq \frac{r\delta}{\ln(\alpha)} \left[ \ln \left( \frac{1 - \beta}{-\beta \ln(\beta)} \right) - \frac{1 - \beta + \beta \ln(\beta)}{1 - \beta} \right]$$

Note that $\beta$ is a function of the loss rate in the last cycle. If we assume the loss rate to be 1, i.e., almost all packets were lost, this gives us $\beta = 0.5$, and we can get a significantly liberal estimation of required queue size as follows, with $G \approx 1.2228$:

$$Q \geq G r \delta$$
\section{Queue Growth Comparison with Shapers}

We noted in previous sections how phantom queues for a policer need to be sized significantly bigger than their shaper counterpart. For example, we saw that to ensure correct rate enforcement for a TCP Reno flow, the phantom queue must be sized $O(BDP^2)$, where BDP is the product of bottleneck rate ($r$) and base propagation delay ($D$), on the other hand a shaper queue only needs to be sized to $O(BDP)$. The root cause of this difference stems from the lack of queuing delay in the case of policers. Generally, we can understand that the standing queue size increases by the difference between the enqueue and dequeue rates:

\begin{equation} \label{eq:dQdt}
\frac{dQ}{dt} = r(t) - r
\end{equation}

For rate-based protocols, it is easier to understand why queue growth rate is higher for a policer versus a shaper. Rate-based CCAs often have a component that also reacts to changes in observed delay, a direct indicator of standing queue size. However, that part of the rate-based CCAs becomes redundant with the policer, where increases in the standing queue size do not inflate delays. The only active component in these protocols when interacting with traffic policers is the one that reacts to losses, e.g., increment part in GCC. As a result, the sending rate of such CCAs grows significantly faster with a policer compared to a shaper. 

Many window-based CCAs do not have a delay-based component; however, ack-clocking still ensures that the rate grows slowly. Note the $\dot{r}(t)$ equation for window-based CCAs in Eq.~\ref{eq:r_dot}. For a policer, it simplified to Eq.~\ref{eq:r_dot2} because $\frac{dR}{dt} = 0$ for policers; however, this is not the case for a shaper, and we note that as the round-trip time increases ($\frac{dR}{dt} > 0$), the change in sending rate slows down. In fact, using the ack-clocking property, we can evaluate the rate of change of the standing queue size as follows. Assuming no cross-traffic on the link, we know that the standing queue size at any time would be given by:

$$Q(t) = \omega(t) - rD$$

This is because the ack-clocking property of window-based protocols ensures that in-flight packets at any time are capped by $\omega(t)$, out of which the bottleneck can service BDP, $rD$, amount of packets, and remaining packets get queued up. Thus, the queue growth is simply the rate of change of the congestion window.

$$\frac{dQ}{dt} = \frac{d\omega}{dt}$$

Thus, in the case of shapers, the queue buildup rate for window-based CCAs is directly controlled by the window update rule. 
As noted earlier, with policers, acknowledgements for packets arrive earlier than the phantom copies of the packet are actually dequeued from the policer's phantom queue.
This means that the ack-clocking property does not hold, and the queue builds up faster than the above relation.
The queue growth rate with policers can be calculated by plugging the $\dot{r}(t)$ function for the CCA in Eq.~\ref{eq:dQdt}. We now show this behavior using different CCAs below. All the results shown ahead are obtained by simulating a shaper/policer bottleneck serviced at a rate of 10 Mbps with a propagation delay of 100 ms, and the queue size is set to BDP for the shaper and as determined by our framework for the policer's phantom queues.

\begin{figure}
\centering
\captionsetup{justification=centering}
\captionsetup[subfigure]{justification=centering}
\centering
\subfloat[Congestion Window]{\includegraphics[width=0.24\textwidth]{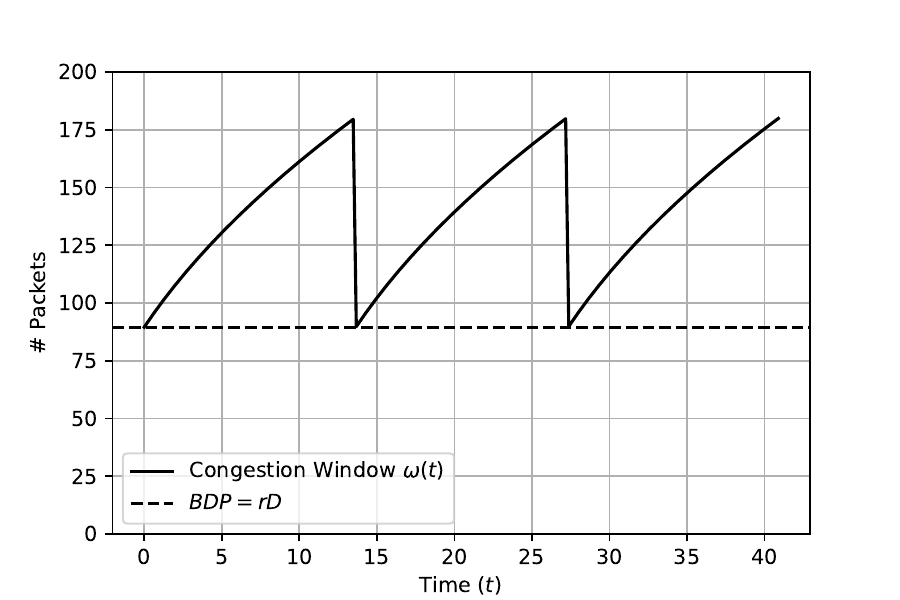} \label{fig:shaper-reno-cwnd}}
\subfloat[Round-trip Time]{\includegraphics[width=0.24\textwidth]{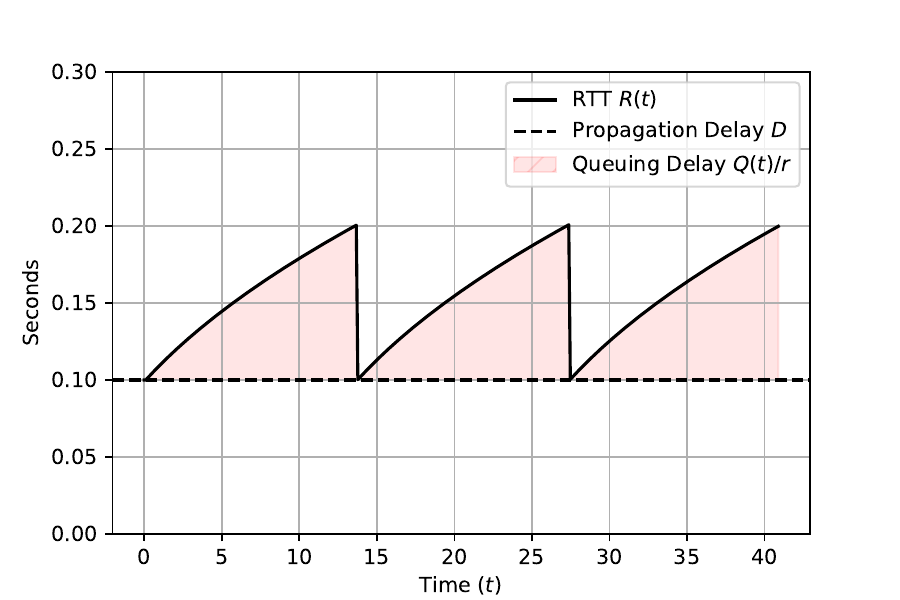} \label{fig:shaper-reno-rtt}}
\subfloat[Sending Rate]{\includegraphics[width=0.24\textwidth]{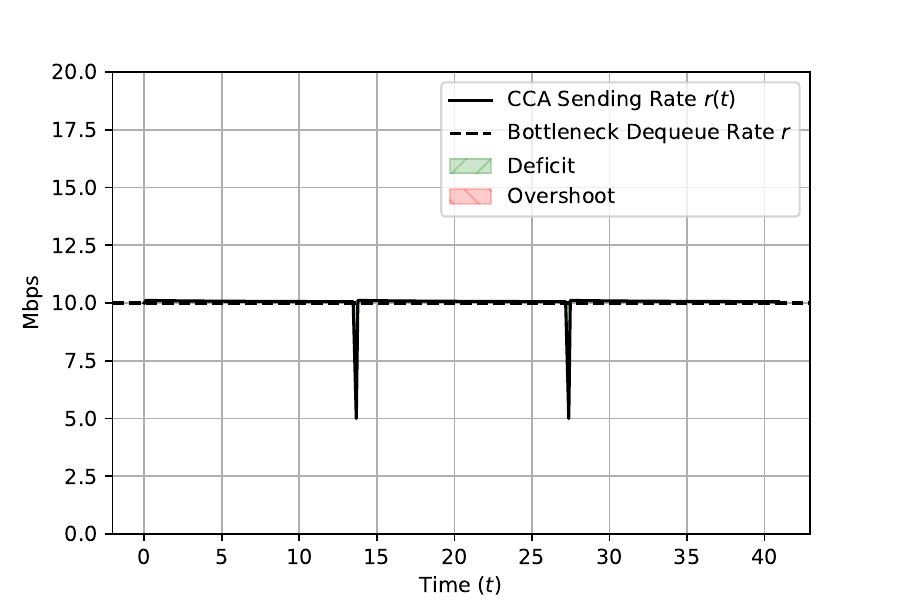} \label{fig:shaper-reno-rate}}
\subfloat[Standing Queue]{\includegraphics[width=0.24\textwidth]{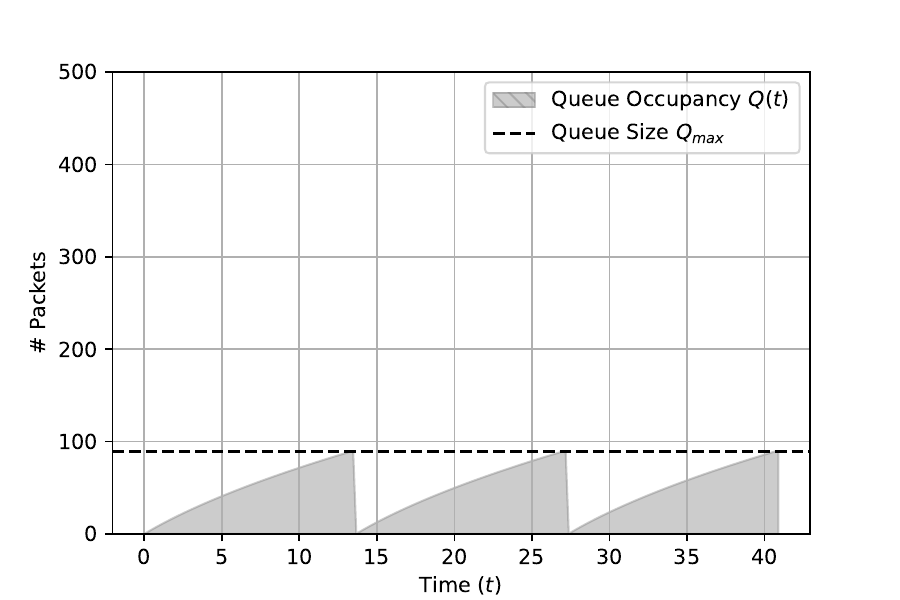} \label{fig:shaper-reno-queue}}
\caption{Interaction of a TCP Reno flow with a shaper-based bottleneck.}
\label{fig:shaper-reno}

\end{figure}

\begin{figure}
\centering
\captionsetup{justification=centering}
\captionsetup[subfigure]{justification=centering}
\centering
\subfloat[Congestion Window]{\includegraphics[width=0.24\textwidth]{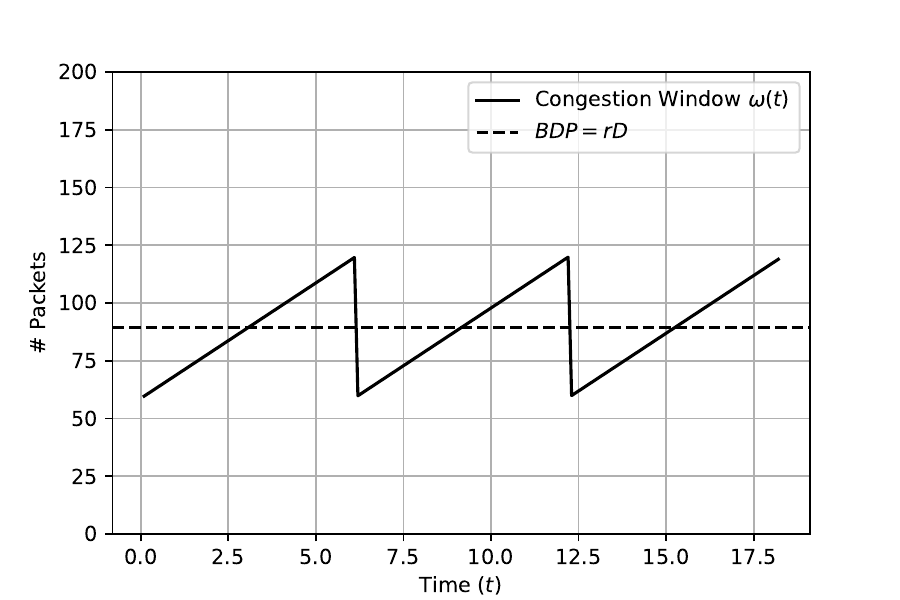} \label{fig:phantom-reno-cwnd}}
\subfloat[Round-trip Time]{\includegraphics[width=0.24\textwidth]{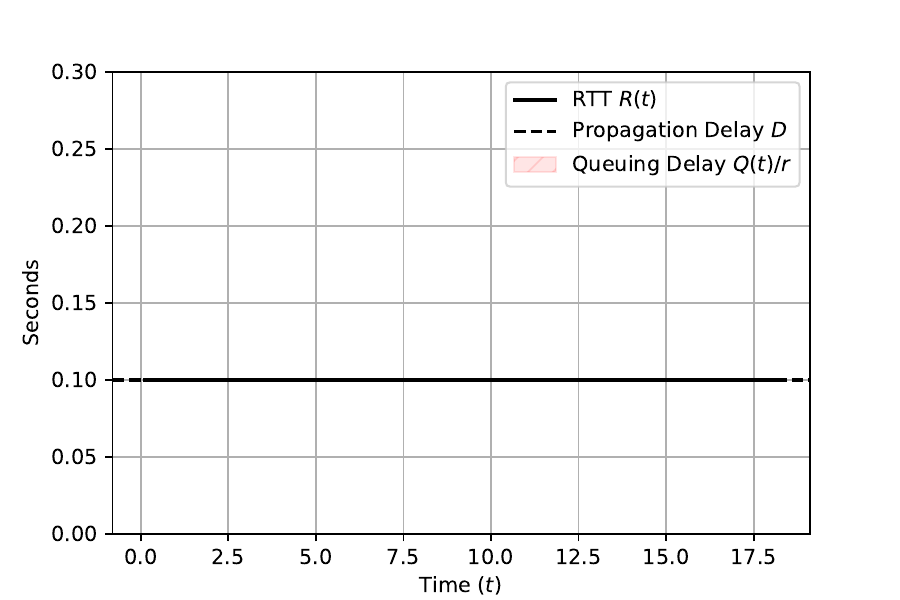} \label{fig:phantom-reno-rtt}}
\subfloat[Sending Rate]{\includegraphics[width=0.24\textwidth]{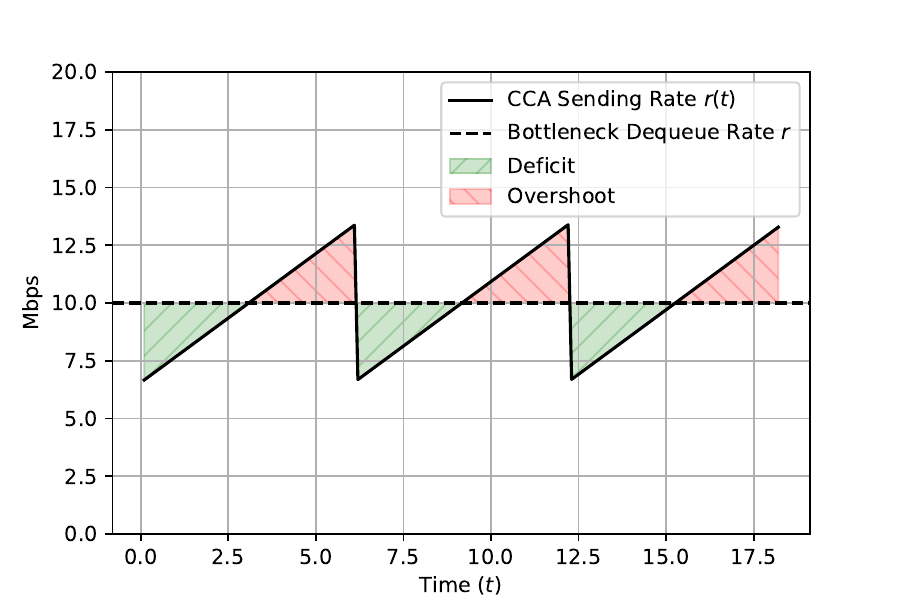} \label{fig:phantom-reno-rate}}
\subfloat[Standing Phantom Queue]{\includegraphics[width=0.24\textwidth]{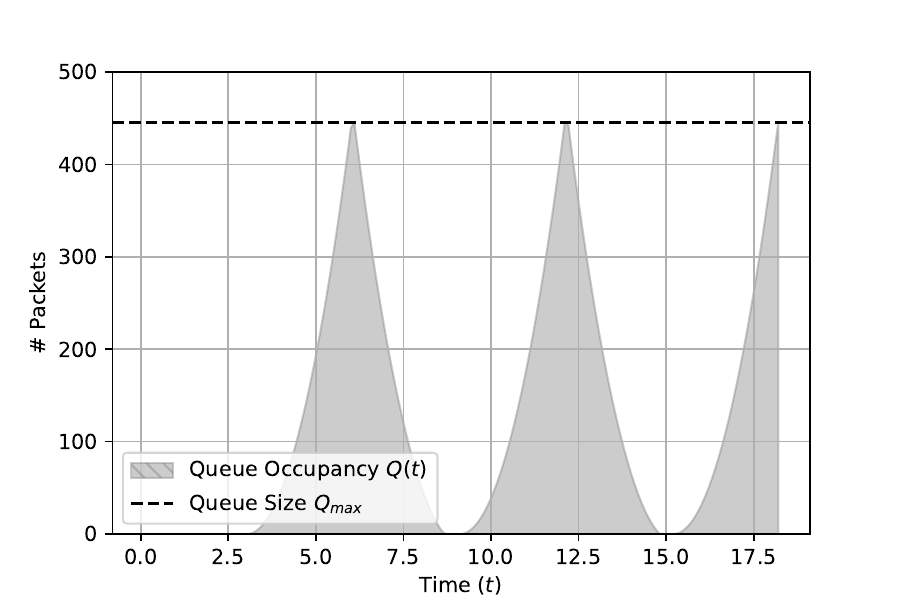} \label{fig:phantom-reno-queue}}
\caption{Interaction of a TCP Reno flow with a policer-based bottleneck.}
\label{fig:phantom-reno}

\end{figure}

\textbf{TCP Reno:}
We show how a TCP Reno flow interacts with a shaper-based bottleneck versus a policer-based bottleneck. Particularly, we show how the congestion window, round-trip time, sending rate, and standing queue size evolve over time in Figure~\ref{fig:shaper-reno} for a shaper and in Figure~\ref{fig:phantom-reno} for a phantom queue policer. First, we see an almost identical pattern when it comes to how the congestion window is updated over time, ignoring the minimum and maximum congestion window. The key difference we expect to see is in end-to-end round-trip times, which, as we see, increases as the queue builds up for the shaper, but not for the policer. This difference significantly affects the sending rate, which can be computed by dividing $\omega(t)$ by $R(t)$. Here, we see that the rate varies significantly for policer. This is because RTT remains constant for a policer, thus any changes in congestion window directly translate into the same magnitude change in sending rate. With shaper, an increase in congestion window is accompanied by a proportional increase in RTT; the sending rate, thus, remains pretty close to the bottleneck rate. This is why the standing queue remains pretty low for the shaper, but grows significantly higher for the policer.

\begin{figure}
\centering
\captionsetup{justification=centering}
\captionsetup[subfigure]{justification=centering}
\centering
\subfloat[Congestion Window]{\includegraphics[width=0.24\textwidth]{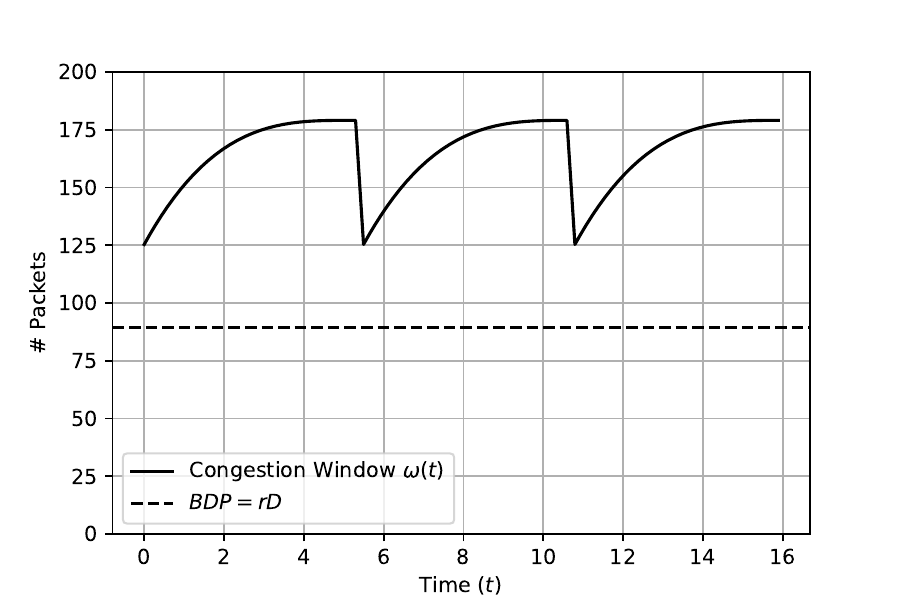} \label{fig:shaper-cubic-cwnd}}
\subfloat[Round-trip Time]{\includegraphics[width=0.24\textwidth]{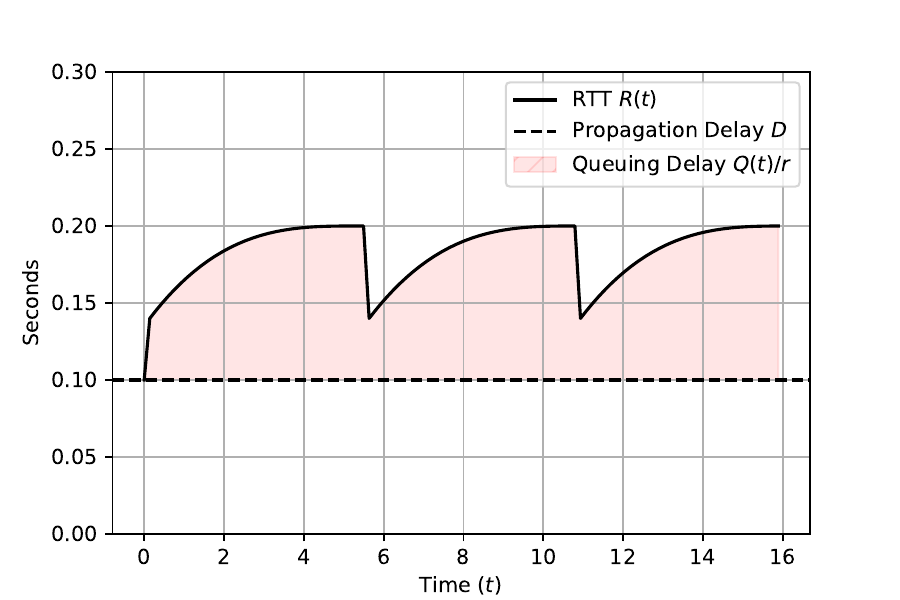} \label{fig:shaper-cubic-rtt}}
\subfloat[Sending Rate]{\includegraphics[width=0.24\textwidth]{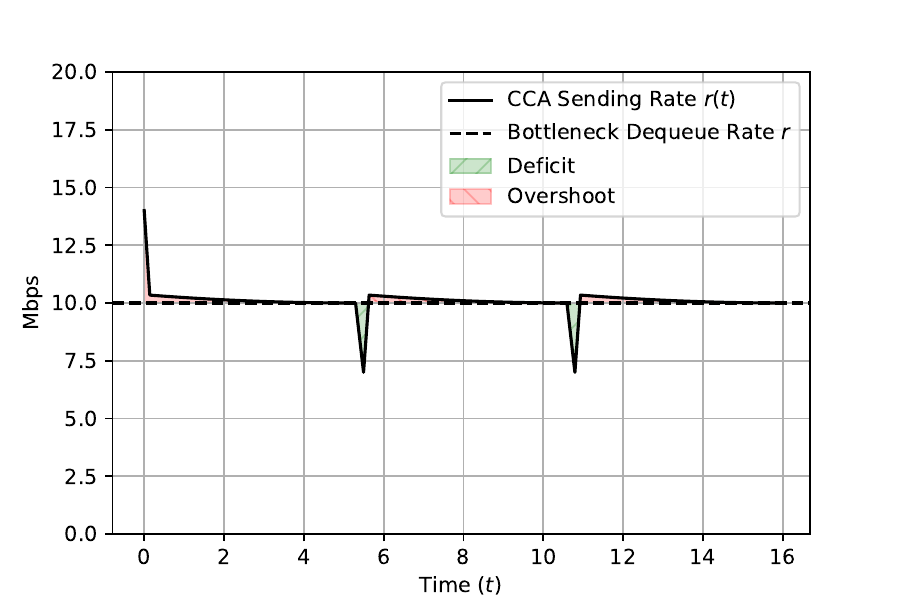} \label{fig:shaper-cubic-rate}}
\subfloat[Standing Queue]{\includegraphics[width=0.24\textwidth]{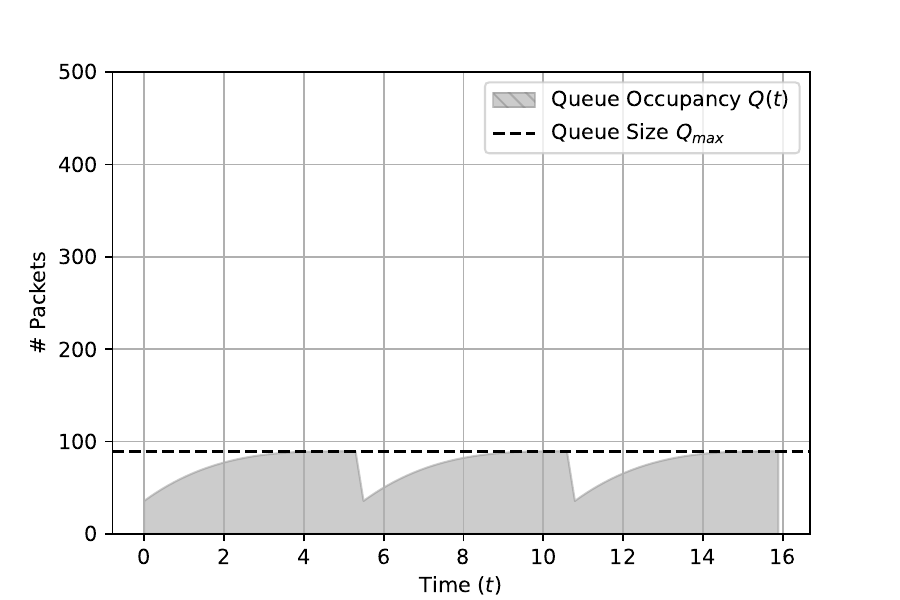} \label{fig:shaper-cubic-queue}}
\caption{Interaction of a TCP Cubic flow with a shaper-based bottleneck.}
\label{fig:shaper-cubic}

\end{figure}

\begin{figure}
\centering
\captionsetup{justification=centering}
\captionsetup[subfigure]{justification=centering}
\centering
\subfloat[Congestion Window]{\includegraphics[width=0.24\textwidth]{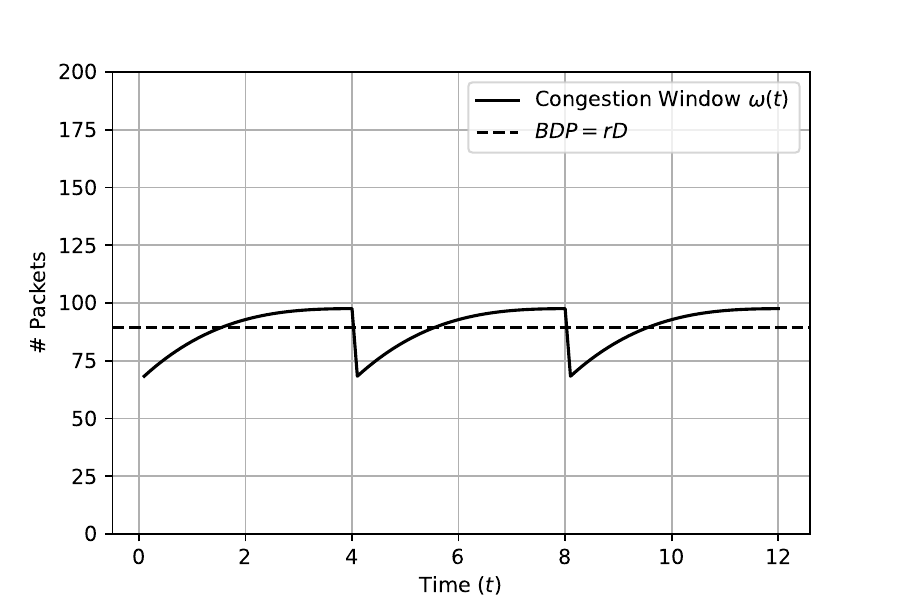} \label{fig:phantom-cubic-cwnd}}
\subfloat[Round-trip Time]{\includegraphics[width=0.24\textwidth]{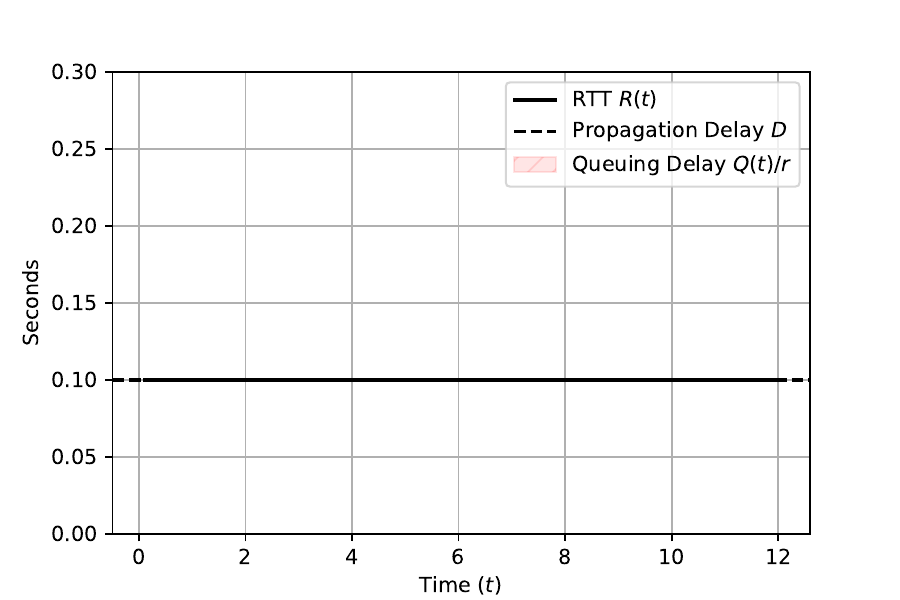} \label{fig:phantom-cubic-rtt}}
\subfloat[Sending Rate]{\includegraphics[width=0.24\textwidth]{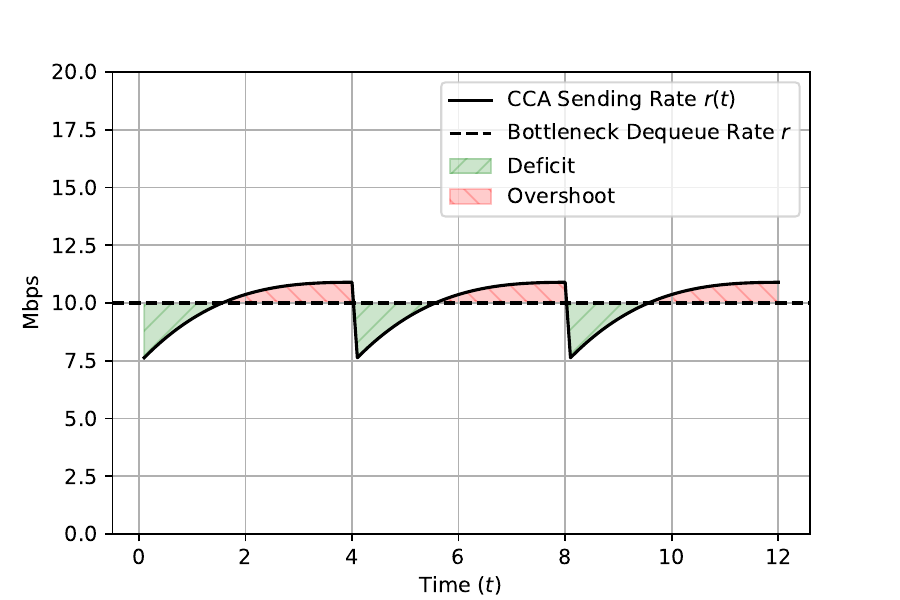} \label{fig:phantom-cubic-rate}}
\subfloat[Standing Phantom Queue]{\includegraphics[width=0.24\textwidth]{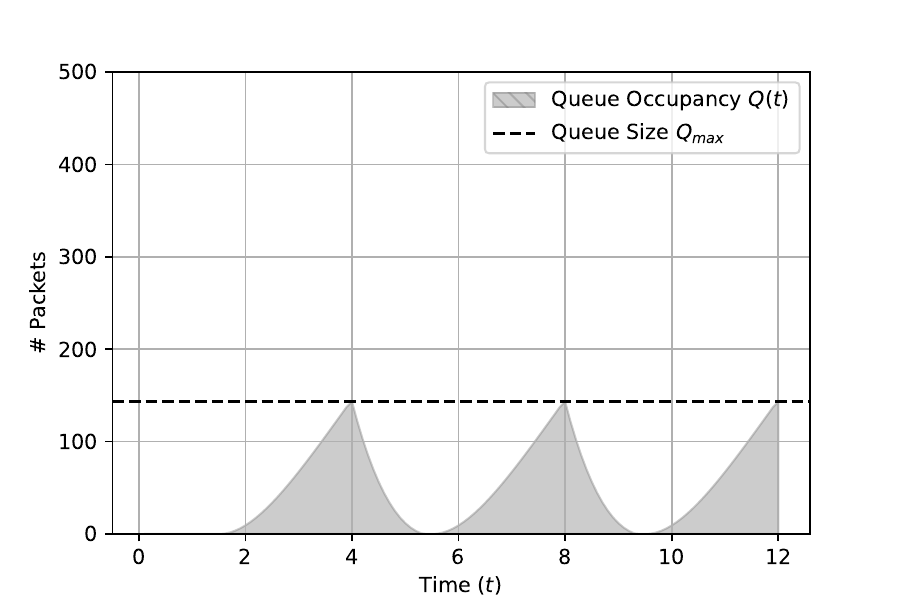} \label{fig:phantom-cubic-queue}}
\caption{Interaction of a TCP Cubic flow with a policer-based bottleneck.}
\label{fig:phantom-cubic}

\end{figure}

\textbf{TCP Cubic:}
We now show the same set of results for TCP Cubic's interaction with a shaper in Figure~\ref{fig:shaper-cubic} and with a phantom queue policer in Figure~\ref{fig:phantom-cubic}. Cubic's congestion window does not increase each round-trip linearly. Instead it grows faster at first and then slowly based on the time elapsed since the last window reduction; therefore, with shaper we can see a relatively significant change in sending rate. With policer, we see identical behavior as Reno, where the sending rate varies significantly more compared with the shaper, accumulating a standing queue at a fast rate.

\bibliographystyle{ACM-Reference-Format}
\bibliography{references}

\end{document}